\documentclass[aps,prl,amsmath,groupedaddress,letterpaper,showpacs,twocolumn]{revtex4}
\usepackage{graphicx,array,tabularx,epsfig,color,xcolor}
\usepackage{epstopdf}
\begin{document}

\title{Are pinholes the cause of excess current in superconducting tunnel junctions? A study of Andreev current in highly resistive junctions}
\author{Tine Greibe}
\email[]{tine@greibe.dk}
\author{Markku P. V. Stenberg}
\email[]{markku.stenberg@iki.fi}
\author{C. M. Wilson}
\author{Thilo Bauch}
\author{Vitaly S. Shumeiko}
\author{Per Delsing}
\affiliation{Department of Microelectronics and Nanoscience,
  Chalmers University of Technology, SE-41296 G\"oteborg, Sweden}
\begin{abstract}
In highly resistive  superconducting tunnel junctions, excess subgap current is usually observed and is often attributed to
microscopic ÒpinholesÓ in the tunnel barrier. We have studied the subgap current in superconductor–--insulator--–superconductor (SIS) and superconductor–--insulator–--normal-metal (SIN)   junctions. In Al/AlO$_x$/Al junctions, we observed a decrease of 2 orders of magnitude in the current upon the transition from 
the SIS to the SIN regime, where it then matched theory. In Al/AlO$_x$/Cu junctions, we also observed generic features of coherent diffusive Andreev transport in a junction with a homogenous barrier.  We use the 
quasiclassical Keldysh-Green function theory to quantify single- and two-particle tunneling and find good agreement over 2 orders of magnitude in transparency.  We argue that our observations 
rule out pinholes as the origin of the excess current. 
\end{abstract}
\date{\today}
\maketitle
Superconducting tunnel junctions have become increasingly important devices in applications ranging from medical and astrophysical sensors to quantum computing.  A hallmark of tunnel junctions, key to these applications, is their minimal dissipation.  This dissipation is often parameterized by a subgap conductance $G_{sg}$ in parallel with an ideal tunnel element.
Even though $G_{sg}$ is relatively small at low temperatures, it is often observed to be orders of magnitude larger than what is predicted by theory.  This excess dissipation is emerging as a potential limitation in a host of new 
applications. For instance, 
this may be a source of energy relaxation in superconducting qubits \cite{makhlin01} and in tunable resonators \cite{cqed}. In single-electron 
turnstiles, this leakage may limit the ultimate accuracy of a future current 
standard \cite{pekola07}.

To elucidate the problem, we note that the tunnel model of superconducting junctions predicts that $G_{sg}$ should decrease exponentially as a function of temperature \cite{BaroneBook}. 
This is true for both superconductor--insulator--superconductor (SIS) and 
superconductor--insulator--normal-metal (SIN) junctions. In experiments with 
highly resistive junctions, however, an exponential dependence is observed for 
temperatures down to approximately 10 \% of the critical temperature of the 
superconductor, after which a saturation is observed at a value a few orders of magnitude smaller than the normal-state tunnel conductance, 
$G_N$ \cite{kleinsasser93,gubrud01}.

In spite of being universally observed, the origin of this temperature 
independent subgap 
current has remained a puzzle for decades. One of the accepted explanations is provided by multiparticle tunneling 
\cite{schrieffer63} and multiple Andreev reflections \cite{Klapwijk1982}. This has been successfully used to explain the subgap current in low resistance junctions \cite{kleinsasser95}. However, even from the first observation \cite{taylor63}, 
experiments on highly resistive junctions have revealed a drastic discrepancy between the measurements and 
theoretical predictions. According to theory
\cite{schrieffer63,bratus95,bezuglyi06}, the subgap current in uniform SIS junctions at low temperatures should have a series of current steps at voltages where different multiparticle processes are activated.  In theory, the ratio of the current below and above each step is proportional to the junction transparency, $\Gamma$.
Therefore, in junctions with $\Gamma\sim 10^{-5}- 10^{-6}$ one expects a current step near $eV=2\Delta$ with ratio $10^{-5}- 10^{-6}$.  However, the experimental values typically saturate around $10^{-2} - 10^{-3}$ \cite{taylor63,kleinsasser95,gubrud01,lang03,milliken04,oh05,castellano06}. 
In SIN junctions, one similarly expects the ratio $G_{sg}/G_N$ to  be of order 
$10^{-5} - 10^{-6}$. However, a similar saturation is typically observed
\cite{pothier94,quirion02,lotkhov06,rajauria08} (though a smaller $G_{sg}$ was 
recently reported in \cite{pekola10}). This discrepancy has often been explained 
\cite{schrieffer63} by assuming microscopic defects in the tunnel barrier, commonly known as ``pinholes," which cause a greatly enhanced local transparency ($\Gamma > 10^{-3}$). In fact, a large subgap conductance has been considered an indicator of a poor quality tunnel barrier \cite{pekola07, kleinsasser95, oh05, lotkhov06}.

\begin{figure}
\includegraphics[width=0.45\textwidth]{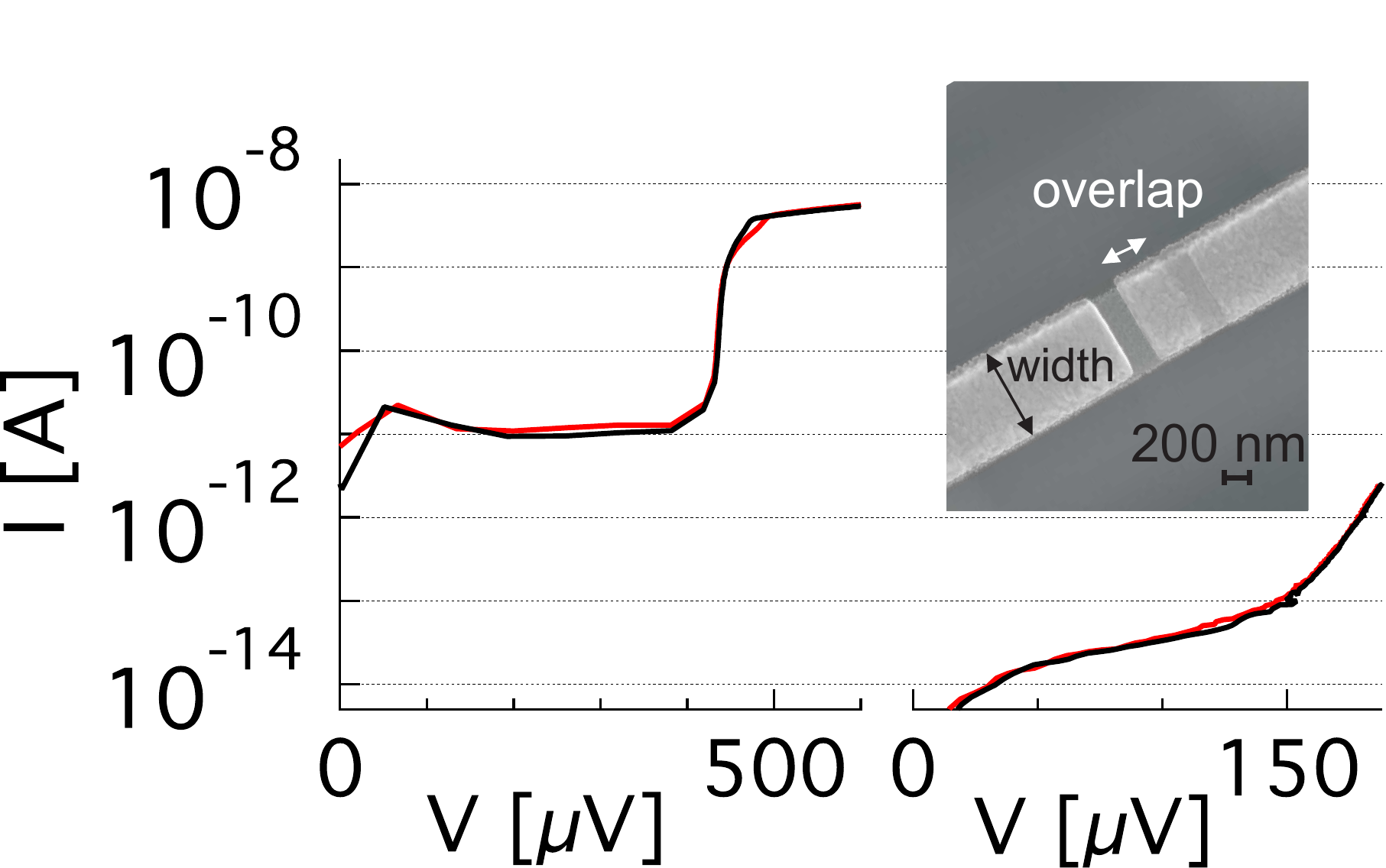}
\caption{\label{SEMfig_IV} (Color online) I-V characteristics of two typical 
low-transparency Al/AlO$_x$/Al junctions (red and black) at $T<100$ mK. 
In the SIS regime (left, $B = 0$ mT), a step is observed with the 
ratio of currents below and above $eV=2\Delta$ being
$ 2 \times 10^{-3}$, consistent with previously published work.
However, 
the subgap current is strongly suppressed upon the transition from the SIS to the SIN regime 
(right, $B > 300$ mT) and approaches the theoretically predicted limit. 
The inset shows an electron micrograph of the junction.}
\end{figure}

\begin{figure} 
\includegraphics[width=0.45\textwidth]{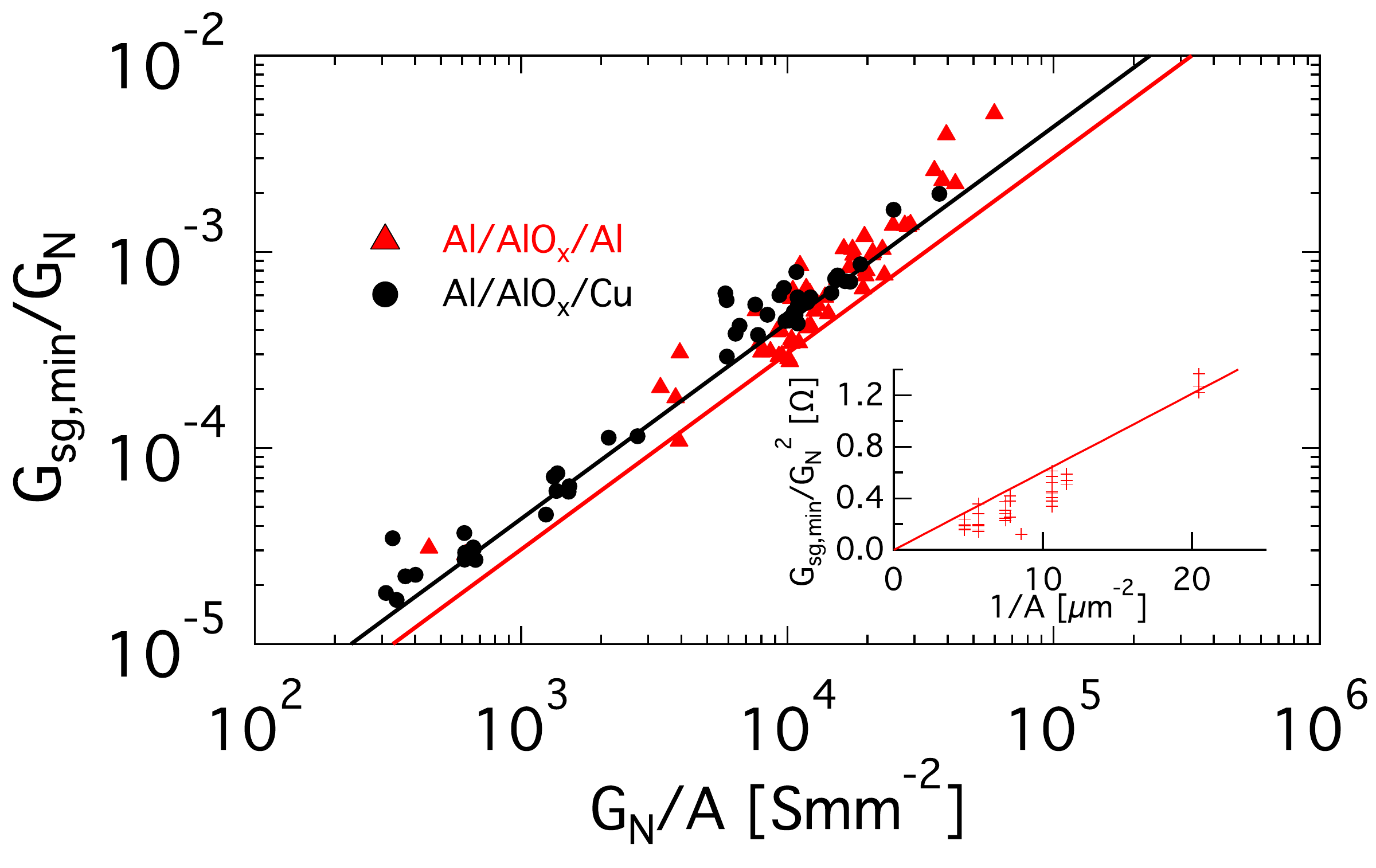} 
\caption{\label{RsgRn_Rn} (Color online) Ratios $G_{sg,min}/G_N$
versus normal conductance per geometric area $G_{N}/A$ in SIN junctions;
experimental points are indicated with dots and triangles, the lines show theoretical predictions of Eq.~(1). Inset: For a fixed $G_N$, we observe that 
$G_{sg,min}$ is inversely proportional to $A$, thus exhibiting a linear 
dependency on $\Gamma$ as suggested by Eq.~(1).} 
\end{figure}

In this Letter, we report on an extensive study of the differential subgap conductance $G_{sg}$ in tunnel junctions
for a wide range of $G_N$. First, we have investigated 
Al/AlO$_x$/Al junctions (Fig.~\ref{SEMfig_IV}), similar to those used in qubit circuits. We investigated both the SIS and SIN transport regimes in these junctions.  This was done by making one electrode thicker and applying an external magnetic field to suppress the superconductivity in this electrode. The two-particle current should not change significantly during this crossover according to theories for diffusive SIN \cite{volkov, hekking94} and SIS  \cite{bezuglyi06} junctions. However, in drastic contradiction to the theoretical expectation, we observed a \textit{decrease} of up to 2 orders of magnitude in the subgap current in the SIN regime (Fig.~\ref{SEMfig_IV}). $G_{sg}/G_N$ then reaches values predicted by the theory of diffusive Andreev transport \cite{volkov, hekking94,bezuglyi06}, exhibiting a linear dependence on $G_N$ over more than 2 orders of magnitude. We ultimately achieve $G_{sg}/G_N \sim 10^{-5}$ (Fig.~\ref{RsgRn_Rn}).

To corroborate that we reached the fundamental limit for $G_{sg}$ set by Andreev reflection, we  fabricated 
and investigated generic SIN Al/AlO$_x$/Cu junctions. Here we achieved 
similarly small values of $G_{sg}$, as shown in Fig.~\ref{RsgRn_Rn}. 
Moreover, we observed a zero-bias conductance 
peak (ZBCP), a fingerprint of coherent diffusive Andreev transport 
\cite{volkov,kastalsky91,wees92,hekking94}. This peak is suppressed by magnetic 
field and thus was not observed in the all-Al junctions. We further measured
a dependence of $G_{sg}$ on the effective electronic mean free path in the electrodes. As we will show, these observations indicate homogeneity of the tunnel barriers on the spatial scale of the mean free path. 

Our junctions were fabricated on oxidized Si substrates by the
standard Dolan evaporation technique with in-situ thermal growth of aluminum
oxide. 
Both metal leads are evaporated resistively at a base pressure of $\sim$10$^{-6}$ mbar with a 0.5-1 nm/s
evaporation rate. The junctions were e-beam patterned using a two or three layer 
resist system. 
The Al/AlO$_x$/Al junctions were $\sim$100-400 nm wide with an
overlap of $\sim$400 nm [Fig.~\ref{SEMfig_IV}(a)]. 
A variation of the oxygen dose (the product of pressure and time) from 17 
\mbox{mbar $\cdot$ s} to 8 \mbox{bar $\cdot$ s} yielded R$_{N}$ 
values from 0.2 to 13 k$\Omega$. 
The Al/AlO$_x$/Cu junctions were $\sim$80 nm wide with an overlap of  $\sim$220 nm. A variation of the dose from  
1.9 to 120 \mbox{bar $\cdot$ s} yielded R$_{N}$ values from 1.5 to 200 k$\Omega$.  We note that the growth conditions for the base electrode and barrier are the same in both types of devices.  We therefore assume that the barrier qualities are similar in both.  
The junctions were measured in a dilution 
refrigerator at temperatures below 100 mK. Each dc line was equipped with a 
two-stage RLC filter, a powder filter, and with 
2 m of thermocoax cable. The junctions were voltage biased and the current 
was read out either through a bias resistor or a transimpedance amplifier. The 
former method introduced excess noise for the high-R$_{N}$ junctions which is
why the latter method was used for these junctions.

To understand the experimental data, we applied a theory of the superconducting 
proximity effect based on quasiclassical Keldysh-Green function techniques 
\cite{volkov, hekking94}. Both single- and two-particle processes contribute to the subgap conductance. The single-particle contribution is given by the conventional tunnel model \cite{BaroneBook}.  The two-particle Andreev conductance is given by
\begin{eqnarray}
\label{eq:G_A}
&&G_{A}(V) = G_N {3\Gamma\over 4 l}  \int_{0}^{\Delta} dE \, 
\frac{\Delta}{\sqrt{\Delta^2-E^2}}\nonumber\\
&&\times{\rm Re}\left[ \sqrt{\frac{D}{2iE}}\tanh\sqrt{E\over 2iE_{Th}}\right] 
\partial_V f_{N}(E,V).
\end{eqnarray}
Here $f_{N}(E,V)=(1/2)\{\tanh[(E+eV)/2k_B T] - \tanh[(E-eV)/2k_B T]\}$ and $E_{Th}=\hbar D/L^2$ is the Thouless energy with $D$  the diffusion constant, $l$  the mean free path, and  $L$ the length of the normal electrode.
Eqn.\ (\ref{eq:G_A}) is valid in the experimentally relevant limit of phase-coherent diffusive transport over the distance $L\gg l$ and assuming that $G_N$ is significantly smaller than the conductance of the 
normal lead. Furthermore, it is assumed that $E_{Th}\ll\Delta $.

Important features of the Andreev conductance (\ref{eq:G_A}) which are relevant for interpretation of the experiment are i) the presence of an additional factor $\Gamma$ indicating the two-particle origin of the Andreev transport, ii) the enhanced value at zero voltage, $G_A(0)=G_N\Gamma (3 L/4 l)$ (ZBCP), and iii) the dependence on $l$.

The ZBCP is indicative of the coherent, diffusive transport regime.  It is explained by an electron-hole transmission resonance formed by the interplay between Andreev reflection and scattering by impurities \cite{wees92,hekking94}. The resonance is destroyed by an external magnetic field $B$, and the ZBCP disappears according to \mbox{$G_{A}(V=0, B)=$} \mbox{$G_{A}(V=0) \tanh(b)/b$}, where $b = \sqrt{2}\lambda_{L} L e B/\hbar$, and $\lambda_{L}$ is the London penetration length \cite{volkov}. The ZBCP is usually observed in high-transparency SIN junctions based on two-dimensional electron gases \cite{kastalsky91}, though it has also been found in metallic junctions with intermediate transparencies 
\cite{pothier94,rajauria08}. Other transport characteristics of diffusive SIN 
junctions, e.g., shot noise, also exhibit a zero-bias anomaly \cite{stenberg02}.

Figure \ref{RsgRn_Rn} shows the measured ratios $G_{sg,min}/G_N$ 
for a wide range of barrier thicknesses in Al/AlO$_x$/Al and Al/AlO$_x$/Cu junctions measured in the SIN regime. Here $G_{sg,min}$ indicates the minimum value of $G_{sg}$
approximately at $eV\approx \Delta/2$. The conductance per unit area, $G_N/A$, where $A$ is the junction area, was varied over more than 2 orders of magnitude. 
The measured ratios $G_{sg,min}/G_N$ 
are proportional to $G_N/A$ as the theory 
(red and black lines) predicts, down to the lowest values measured $\sim$10$^{-5}$. 
At a fixed $G_N$,  we observe that $G_{sg,min}$ is inversely proportional to 
$A$, indicating a linear dependence on $\Gamma$ (inset of Fig.~\ref{RsgRn_Rn}).

\begin{figure} 
\includegraphics[width=0.45\textwidth]{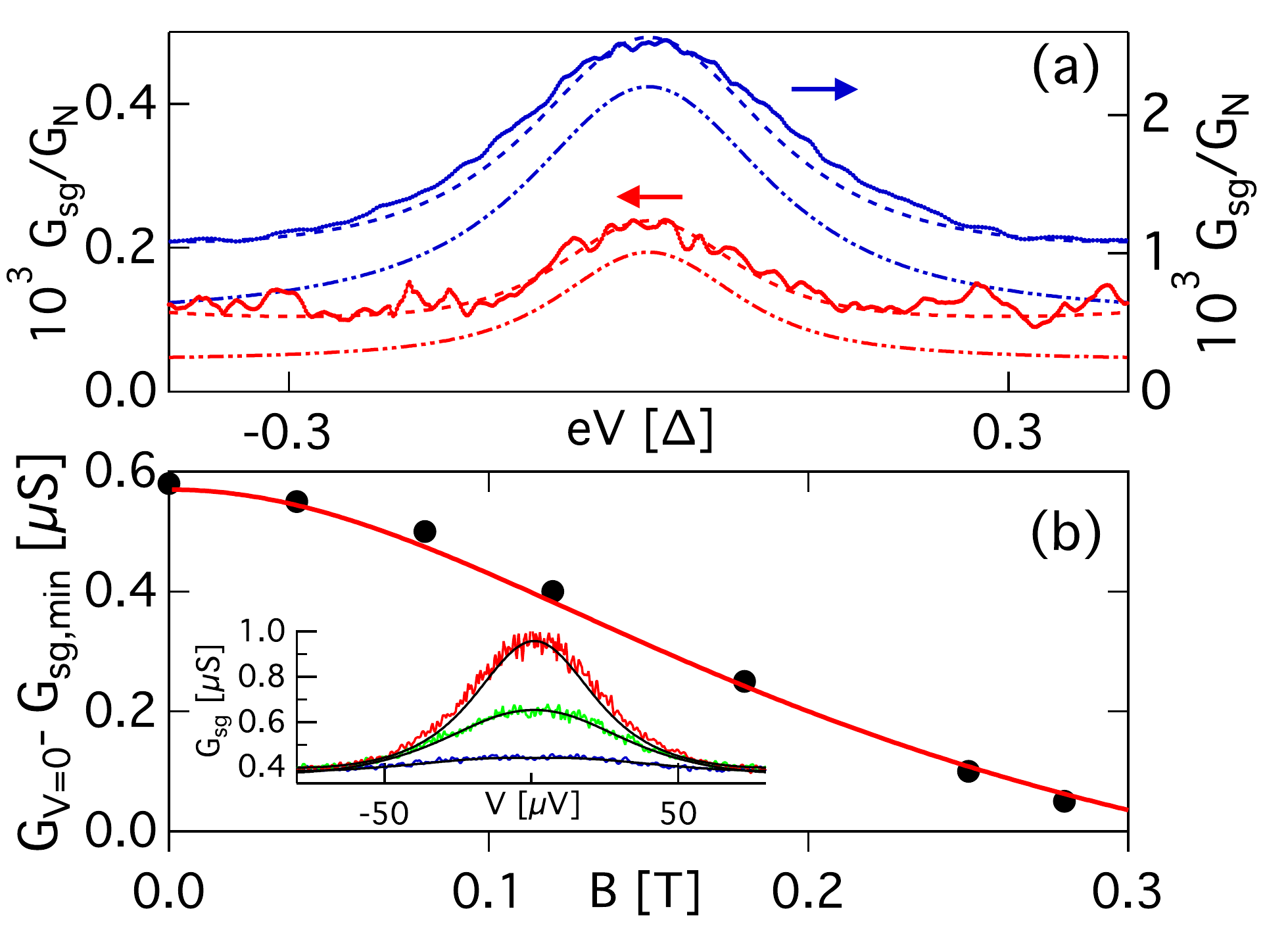} 
\caption{\label{GV} (Color online)
(a) Subgap conductance vs voltage measured in SIN tunnel junctions with
transparencies $\Gamma = 1\times 10^{-4}$ (blue, right y-axis) and 
$\Gamma = 2\times 10^{-5}$ (red, left y-axis), solid lines. The best fit (dashed lines) is obtained by adding the single-particle conductance to $G_A(V)$ of Eq.~(\ref{eq:G_A}) (dash-dotted lines), assuming an electronic temperature of 100 mK, and finite values of the Dynes parameter (defined in the text), $\gamma = 4\times 10^{-5}\Delta$ and $\gamma = 4\times 10^{-4}\Delta$, respectively. (b) Suppression of ZBCP by magnetic field $B$ applied parallel to the substrate. Dots are experimental data. The line is theory. Inset shows $G(V)$ curves for fields: B=0; 180 mT; 280 mT (from top to bottom).}
\end{figure}

$G_{sg}$ vs voltage characteristics have been investigated in detail in all measured  Al/AlO$_{x}$/Cu junctions. Figure~\ref{GV} shows curves for two junctions, with parameters, $G_N = 50\ \mu S, \Gamma = 2\times 10^{-5}$ 
(red solid), and $G_N = 300\ \mu S, \Gamma = 1\times10^{-4}$ (blue solid).  In order to fit $G_N$ and the one- and two- particle contributions to $G_{sg}$ simultaneously, one must make assumptions about the distribution of $\Gamma$ for the many microscopic conduction channels in the junction. We use a minimal model of a \textit{uniform} $\Gamma$ in an \textit{active area} which may be smaller than the geometric area.  We find that the active area is about $13\%$ of 
the geometric area (cf.~\cite{pothier94}).  
The other parameters used for the theoretical fitting are 
$\Delta=210\ \mu$eV, $D_{Cu}=130$ cm$^2/$s, and $L_{Cu}=5\ \mu$m. 
The best fit is achieved assuming an electronic temperature of 100 mK (the cryostat temperatures were 70 and 40 mK, respectively), and a tunneling
density of states that is broadened due to spurious inelastic processes. The broadening is parameterized by adding a small imaginary part $\gamma$ to the quasiparticle energy, $E+i\gamma$ (the Dynes parameter \cite{dynes84,pekola10}), having values of $\gamma = 4\times 10^{-5}\Delta$ and $\gamma = 4\times 10^{-4}\Delta$, 
respectively. Note that $\gamma$ affects the single-particle conductance but not $G_A$.  We found similar values of temperature and $\gamma$ values of the same order for all junctions that were fit.  

We also measured the effect of an applied magnetic field on the 
ZBCP. The results [Fig.~\ref{GV}(b)] are in a good 
quantitative agreement with the theory. The ZBCP disappears in 
magnetic fields larger than 280 mT, even while the Al electrode remains superconducting.

\begin{figure}
\includegraphics[width=0.45\textwidth]{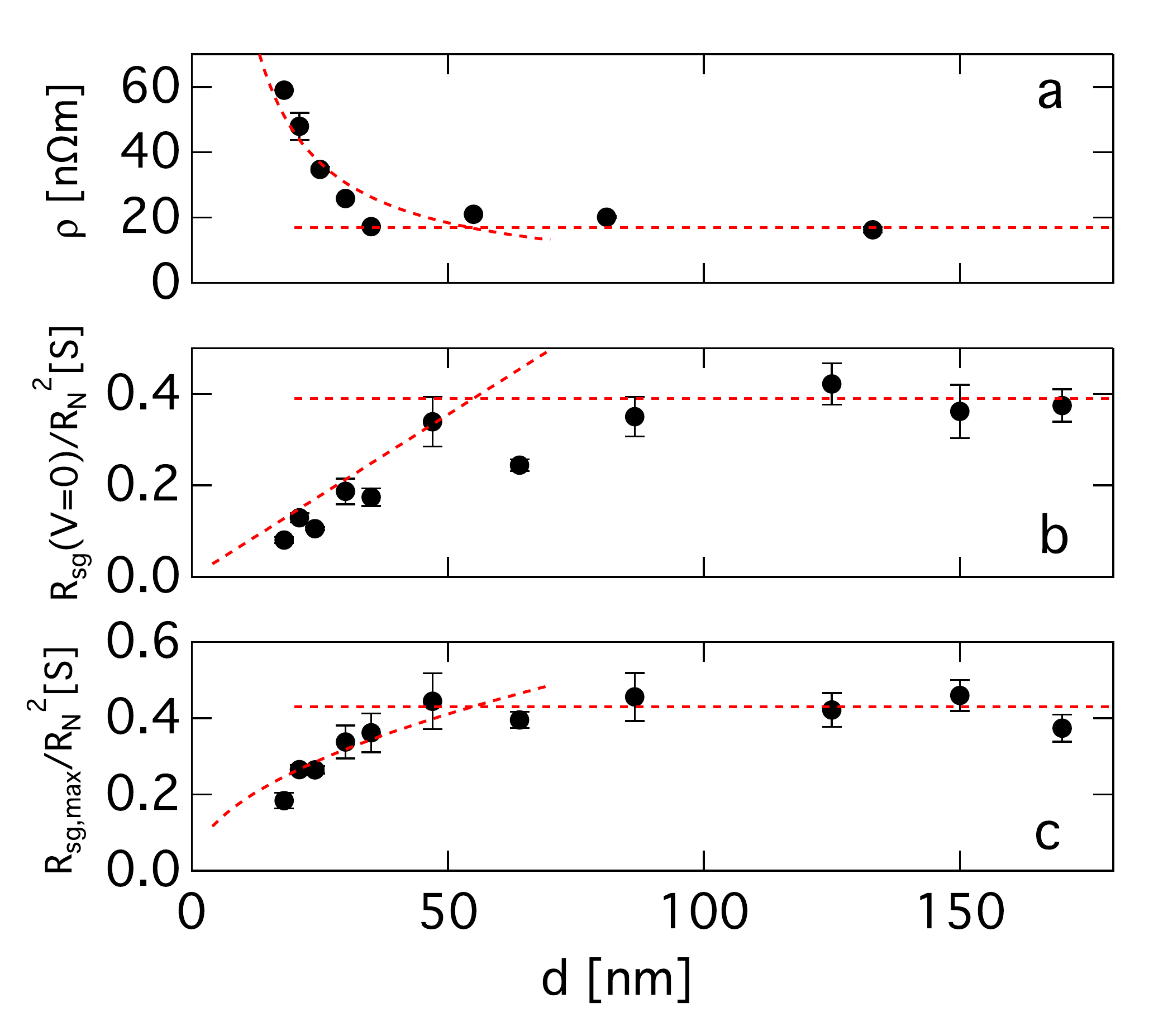}
\caption{\label{thick} (Color online)
Scaling of the transport characteristics with thickness of the Cu electrode 
$d$: (a) specific resistance of the Cu electrode, 
(b) normalized zero-voltage resistance, and (c) normalized 
maximum value of resistance. Dots correspond to averages over several 
measured junctions, the error bars denote the standard deviation. Dashed lines 
show asymptotic behaviors extracted from Eq.~(\ref{eq:G_A}).}
\end{figure}

To verify the dependence of the subgap conductance on the mean free path, we fabricated junctions with different thicknesses 
$d$ of the Cu layer. In bare Cu wires, resistivity as a function of 
$d$ exhibited a crossover from a constant to a $1/d$ dependence at 
$d\approx 50$ nm [Fig.~\ref{thick}(a)], suggesting a crossover to surface dominated electron scattering. 
Figures~\ref{thick}(b) and~\ref{thick}(c) show measurements of $R_{sg}$ vs $d$.  Each point is the average of several junctions.  At approximately the same thickness, $d<50$ nm, the data exhibit a crossover from constant behavior to a variation consistent with Eq.~(\ref{eq:G_A}), namely $R_{sg}(V=0)\propto d$ and $R_{sg,max} \propto \sqrt{d}$.

%
%

From the measured dependence of $G_{sg}$ on electrode thickness, we can draw important conclusions about the uniformity of our tunnel barriers. The employed theoretical model
assumes a homogeneous tunnel barrier on the scale of $l$. The fact that our observations agree with this model implies that any inhomogeneity of the tunnel barrier has a spatial scale larger than 
$\sim$ 50 nm.  If subgap transport were dominated by pinholes with a smaller size, the current would rapidly spread out in the electrode, so that electron scattering would not play a 
role, and transport would resemble that of a ballistic constriction. In that 
case, $G_{sg}$ would not show any dependence on $d$. We note that this 50 nm size scale matches that of the metallic grains in our base electrodes.  This suggests a picture where the tunnel barrier is uniform on any given grain, but varies from grain to grain. The active tunneling area we extract then corresponds to a couple of grains.

Having concluded that the tunnel barriers in our devices do not have pinholes, we must therefore conclude that the greatly enhanced subgap current when one-and-the-same device is measured in the SIS regime cannot be attributed to pinholes.  It must be caused by other mechanisms, \textit{e.g.}, environmental resonances \cite{greibe09}, and it remains an open question.  It is reasonable to extend this conclusion to other junctions fabricated using similar fabrication techniques, which are in fact quite common.

We note that $\gamma$ has implications for decoherence in superconducting qubits. In SIS junctions, 
it gives a residual conductance, $G(V=0)= G_N\gamma/\Delta$, which should lead to relaxation. 
Considering, \textit{e.g.}, a transmon qubit, we obtain
a relaxation time $T_1=\pi\Delta^2/\hbar \omega^2\gamma$. Observed values of $T_1 \sim 6.5\ \mu$s \cite{houck08} would imply $\gamma/\Delta \sim 10^{-4}$, similar to our results, although this is not proof of causation.

In conclusion, we have demonstrated a decrease of subgap current by 2 
orders of magnitude in tunnel junctions as one of the superconducting 
electrodes is made normal. Good quantitative agreement with theory was observed in SIN junctions over a span of more than 2 orders of magnitude 
of the junction transparencies, $\Gamma\sim 10^{-4} - 10^{-6}$, with a minimum value of 
$G_{sg}/G_N \sim 10^{-5}$. We observed all the generic features of coherent diffusive Andreev transport. Taken together, these observations strongly suggest that highly transparent, microscopic pinholes in the tunnel barrier are not the explanation for the observed excess subgap current in highly resistive SIS tunnel junctions. 

This research was partly funded by the Office of the Director of National 
Intelligence (ODNI), Intelligence Advanced Research Projects Activity (IARPA), 
through the Army Research Office.  All statements of fact, opinion or 
conclusions contained herein are those of the authors and should not be 
construed as representing the official views or policies of IARPA, the ODNI, 
or the U.S. Government. We also acknowledge the Swedish Research Council (VR) and the Wallenberg
Foundation for support.

\end{document}